\documentclass[twocolumn]{webofc}
\usepackage[varg]{txfonts} 
\usepackage{booktabs}
\usepackage[table,xcdraw]{xcolor}
\usepackage{url}

\usepackage{textcomp}
\usepackage{breakurl}
\usepackage[breaklinks]{hyperref}
\usepackage{array} 
\newcolumntype{L}[1]{>{\raggedright\let\newline\\\arraybackslash\hspace{0pt}}m{#1}}
\newcolumntype{C}[1]{>{\centering\let\newline\\\arraybackslash\hspace{0pt}}m{#1}}
\newcolumntype{R}[1]{>{\raggedleft\let\newline\\\arraybackslash\hspace{0pt}}m{#1}}

\graphicspath{{graphics/}{graphics/arch/}{Graphics/}{./}} 
\begin{document}
\title{Multi-channel online discourse as an indicator for Bitcoin price and volume}
%
%

\author{\firstname{Marvin Aron} \lastname{Kennis}} 

\institute{Vrije Universiteit Amsterdam, De Boelelaan 1105, 1081 HV Amsterdam, The Netherlands
          }

\abstract{%
This research aims to identify how Bitcoin-related news publications and online discourse are expressed in Bitcoin exchange movements of price and volume. Being inherently digital, all Bitcoin-related fundamental data (from exchanges, as well as transactional data directly from the blockchain) is available online, something that is not true for traditional businesses or currencies traded on exchanges. This makes Bitcoin an interesting subject for such research, as it enables the mapping of sentiment to fundamental events that might otherwise be inaccessible. Furthermore, Bitcoin discussion largely takes place on online forums and chat channels. In stock trading, the value of sentiment data in trading decisions has been demonstrated numerous times~\cite{Bollen2011}~\cite{oh2011investigating}~\cite{smailovic2013predictive}, and this research aims to determine whether there is value in such data for Bitcoin trading models. To achieve this, data over the year 2015 has been collected from Bitcointalk.org, (the biggest Bitcoin forum in post volume), established news sources such as Bloomberg and the Wall Street Journal, the complete /r/btc and /r/Bitcoin subreddits, and the bitcoin-otc and bitcoin-dev IRC channels.
\\

By analyzing this data on sentiment and volume, we find weak to moderate correlations between forum, news, and Reddit sentiment and movements in price and volume from 1 to 5 days after the sentiment was expressed. A Granger causality test confirms the predictive causality of the sentiment on the daily percentage price and volume movements, and at the same time underscores the predictive causality of market movements on sentiment expressions in online communities.  
}
\maketitle

\section{Introduction}\label{sec:introduction}
After the internet forever changed the way investors trade, analyze, and acquire information~\cite{zhang2010day}, it is now giving rise to a new era of financial innovation through the introduction of cryptocurrencies. The expanding market of cryptocurrencies involves capital exceeding USD10 billion as of November 2016~\cite{coindeskmarket}, providing an unusual opportunity to study the emergence of value from currency created solely in the digital realm. Similar to traditional currencies, the value of cryptocurrencies is largely based on supply and demand, driven by the community’s belief in the merit of these currencies. Although hundreds of different cryptocurrencies are now available, Bitcoin is considered to be the the virtual currency that set off the cryptocurrency revolution and now enjoys the largest market capitalization~\cite{coinmarketcap}. 

Research on the influence of online discourse on the Bitcoin exchange price has been scarce, and mostly focused on data from a single channel.~\cite{Kaminski2014} demonstrates that for the prediction of Bitcoin price movements online discourse on Twitter seems to lag behind events occurring on the exchange, rather mirroring what happens on the metaphorical trading floor than predicting what is going to happen in the future. Contradicting results were achieved by~\cite{garcia2015social}, where the increasing polarity of social signals preceded corresponding movements of the Bitcoin exchange price.~\cite{garcia2015social} also shows that incorporating these signals into algorithmic trading models can lead to profitable trading strategies.
					
~\cite{bukovina2016sentiment} looked into the less rational factors for Bitcoin price formation, and found sentiment to play an important role. Findings by~\cite{ciaian2016economics} suggest that macro-economical financial trends do not significantly influence the Bitcoin price. 

Studies regarding the influence of online discourse on traditional stock and foreign exchanges on the other hand have been plentiful, and provide the starting point for this paper.~\cite{tetlock2007giving} investigates the role of media in the stock market, and found that using daily content from the Wall Street Journal, high media pessimism predicted downward pressure on market prices followed by a reversion to the fundamental value of the security. Unusually high or low pessimism can predict high market trading volume. In~\cite{gidofalvi2001using} the predictive value of news reports on stock prices in the short term is asserted. 
	Besides news, social media and other online sources may be used to extract early indicators for investor sentiment, as demonstrated by~\cite{Bollen2011}, where Twitter data was shown to be a leading indicator for the closing prices of the Dow Jones Industrial Average index (DJIA). Google Trends data has previously been used in~\cite{bordino2012web} to show that this information can provide meaningful indicators for upticks in market volume one or more days after increases in search volume, indicating that online behavior away from the exchanges carries anticipatory value for events occurring on exchanges.~\cite{sabherwal2011internet} also shows that  in easily manipulated market environments, such as those of thinly traded stocks, online discussion boards can be effectively used to manipulate the price, even without the presence of fundamental news. 
  
The added value of sentiment data in stock forecasting algorithms is the main motivator behind the research presented in this paper. Should there be a causal connection between Bitcoin-related sentiment and exchange movements, it can be integrated into trading models that trade on the Bitcoin markets and potentially lead to increased profits. These predictions may also aid in reducing risk, as increases in volume tend to be paired with higher volatility. Volatility in turn is an important metric in managing risk in (algorithmic) trading strategies~\cite{jones1994transactions}. 

\subsection{Contributions of this paper}\label{sec:contributions}

The research presented in this paper investigates the predictive value of online discourse originating from multiple channels on Bitcoin price and volume, by analyzing publication volume and sentiment data from these channels concurrently. It extends the current research in this domain that has demonstrated co-movement relations between social signals and market metrics such as price, volatility and volume on various types of financial exchanges. The hypothesis leading this paper is that this online sentiment data has significant predictive causality in relation to inter-day Bitcoin market movements.

\subsection{Outline}\label{sec:outline}
This paper will first discuss the workings of Bitcoin in section~\ref{sec:how-bitcoin-works} before referring to financial theory to determine parallels between Bitcoin and traditional financial instruments in section~\ref{sec:investor-reactions}. Drawing these parallels allows us to select established financial theories as a starting point, given the relative novelty of cryptocurrencies compared to traditional financial instruments. This is followed by an overview of sentiment analysis and its value in financial markets, as well as a detailed description of relevant sentiment analysis algorithms in section~\ref{sec:sentiment-analysis}. The complete data collection to classification approach is explained in section~\ref{sec:approach}, evaluating each classifier against the collected data sources (~\ref{sec:sentiment-classification}), and finally using the best performing classifier for each respective channel in the sentiment classification of all collected data per channel. Building on this, the process of establishing a correlation and causal relationship by means of a Granger causality test between the recorded sentiment and market data is described in section~\ref{sec:correlation}. 

\section{Background and related work}\label{sec:background}
\subsection{How Bitcoin works}\label{sec:how-bitcoin-works}
Bitcoin is an electronic peer to peer payment system first introduced by Satoshi Nakamoto in 2008 and became functionally implemented by 2009~\cite{Nakamoto2008}. Bitcoin transactions are stored on a publicly accessible ledger called the blockchain, allowing everyone to verify, validate and execute transactions that are recorded in this ledger. Instead of having a single authority rule over a currency - such as is the case with fiat currencies and preceding digital currencies, it is now regulated by a decentralized network. Bitcoin's innovation lies in it being the first digital currency to eliminate the need for a central trusted party. It does so by relying on the HashCash proof of work function for transaction processing and verification~\cite{back2002hashcash}. This proof of work requires participating computers in the network (called Bitcoin miners) to conduct increasingly difficult computations. Specifically, miners are required to produce an SHA-256 hash with an arbitrary number of leading zeros from the transaction block. Upon solving those computations they are awarded with newly minted Bitcoin (providing an incentive to solve the computations), and the transactions in the block for which the computation was solved become verified. While these proof of works are complex to compute, they are easy to verify~\cite{back2002hashcash}.  

With a market cap that surpassed USD10 billion in November 2016 according to Coindesk~\cite{coindeskmarket}, Bitcoin can now be considered more than an idealistic hobby project rooted in ideas put forward by the cypherpunk community~\cite{back2002hashcash}. It has attracted attention from investors, consumers, criminals, and governments worldwide~\cite{popper2015digital}. However still not adopted by the masses as a day-to-day currency, adoption of the cryptocurrency has moved past underground communities and mere hobbyists. The cryptocurrency is now accepted even by larger corporations such as Dell~\cite{dellbitcoin} and Overstock~\cite{overstockbitcoin}, raising the question whether Bitcoin can be considered a real currency as consumers can increasingly use it for their traditional purchases. 

\subsection{Investor reactions to news}\label{sec:investor-reactions}
Although primarily intended to be used as a currency by its inventor, Bitcoin should still be seen as a speculative investment according~\cite{Yermack2013} and~\cite{ciaian2016economics}. A proper currency acts as a medium of exchange, a store of value, and a unit of account~\cite{Yermack2013}. Bitcoin largely fails to satisfy these criteria. Firstly, it is still not widely accepted compared to established payment methods and as such can not be considered an effective medium of exchange. Next to that, the Bitcoin exchange price is highly volatile, making it a poor store of value. In the absence of large institutional investors, speculative investments in Bitcoin are more likely driven by retail or individual investors called noise traders. According to behavioral finance research by~\cite{kumar2006retail} and~\cite{barber2011behavior}, noise traders are prone to exhibit less rational market behavior. The sentiment as it may be expressed online could therefore serve as an interesting  indicator for exchange movements, as these noise traders might be influenced by them. In the context of Bitcoin, online messages can disclose new or previously private information that fundamentally alters Bitcoin valuations, such as when new merchants accept Bitcoin or forthcoming regulations limit its use, but may also disclose attempts of market manipulation in the absence of fundamental news. 

\begin{figure}
  \centering
  \includegraphics[width=\columnwidth]{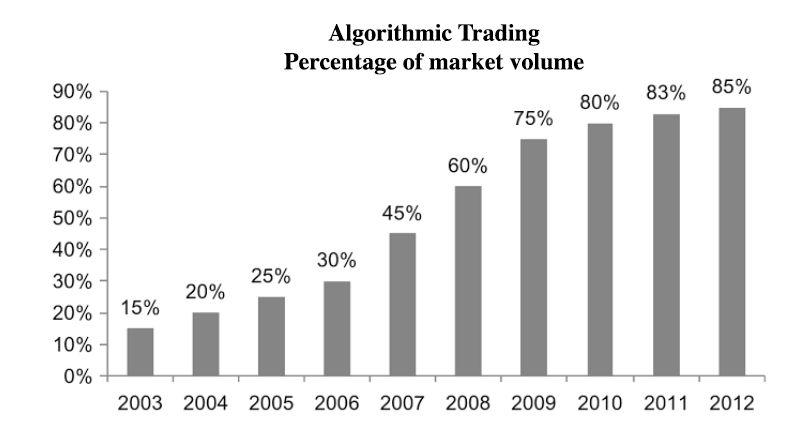}
  \caption{Increase of algorithmic trading as a percentage of market volume from 2003 to 2012~\cite{glantz2013multi}}.
  \label{fig:algoincrease}
\end{figure}

Research concerning the influence of news reports on stock market returns has been abundant.~\cite{pearce1984stock} has shown that elements of surprise in news reports are a strong indicator for stock price movements shortly after the surprising news has been published. This is in line with the (widely disputed) efficient market hypothesis postulated by Fama in 1970~\cite{fama1970efficient}, which states that all available information regarding a security is already reflected in the stock price. If the efficient market hypothesis were to be true, it would prevent anyone from exploiting securities that are mis-priced because large movements are only caused by unforeseen events, and not by misinterpretation of the true value of the security by other market participants (overvaluing and undervaluing participants will cancel each other out on average). This implies that only if a news report contains unexpected or previously unknown information, it will move the exchange price. In the meantime, the price will follow a random-walk. However, the sole presence of high-frequency sentiment analysis on financial exchanges discredits the efficient market hypothesis on short-term intervals~\cite{kleinalgo}. This is where the value of computerized sentiment analysis becomes apparent; market participants reacting rapidly to new insights (whether this insight stems from opinion mining or elsewhere) will have a higher likelihood of being able to profit from it before the information reaches other parties. The rapid rise of high-frequency trading (HFT) firms over the past decade (trading in nanosecond time frames), precipitated by the introduction of large-scale algorithmic trading on stock exchanges attest to this~\cite{chaboud2014rise}. Figure~\ref{fig:algoincrease} shows the increase in algorithmically  executed trade orders between 2003 and 2012. 

~\cite{krawciw} estimates that up to 40\% of equity and 15\% of foreign exchange trades were initiated by high-frequency traders in 2016. News-based trading is an established practice in the HFT industry.  An illustrative example of the convergence of sentiment analysis and algorithmic high-frequency trading is the `Hash Crash' that took place on April 23, 2013, in which a hacked Twitter account of the Associated Press spread false rumors about an attack on the White House, subsequently causing a drop in the Dow Jones Industrial Average of 143 points in minutes~\cite{fthash2013}. Besides the need for speed, unstructured data is being generated in such vast amounts over such diverse channels, that no single human will be able to find, let alone process, all of this. It then follows that different market participants will act upon the relevant information in varying speeds. 

According to principles of behavioral economics, opinions which are not necessarily driven by the release of topic-related news are also thought to influence movements in financial markets, as investors are susceptible to cognitive bias and other predictable human errors~\cite{barber2011behavior}. It therefore seems valuable to examine the overall sentiment in Bitcoin-related communities and Bitcoin-related news in parallel. Bollen et al. (2011) investigate the use of Twitter mood as a forecasting mechanism for the Dow-Jones Industrial Average index (DJIA) in~\cite{Bollen2011}, without filtering on communities and find an accuracy of 87.7\% in predicting daily up and down movements in the closing values of the index. Due to Bitcoin's low penetration in the consumer market and comparatively small ecosystem, it is unreliable to study the mood of a representative sample of the general population as they likely do not interact with Bitcoin on a daily basis and might consequently not represent the opinions of the individuals that do actively interact with Bitcoin. The analysis to be carried out later in this paper will thus be limited to Bitcoin communities as they can be found on online forums and chat channels.

Further, financial news may concern a single company, but news not directly written about a single company or index might also influence its price. Take for instance an announcement about a cut in oil production, which might influence stocks of industries and companies dependent on this resource. The same holds true for Bitcoin, where changes in international monetary policy might, for example, make Bitcoin a more attractive refuge, increasing attempts to regulate the currency will likely have a negative effect on the demand for the cryptocurrency. This type of secondary news is out of the scope of this work.

\subsection{Sentiment Analysis}\label{sec:sentiment-analysis}
Opinions play an important role in human decision making processes, and individuals are increasingly turning to the (social) web to aid them in making those decisions~\cite{pang2008opinion}. Relevant data is fragmented across multiple sources or buried in lengthy forum discussions and blog posts. As the amount of unstructured data on the web keeps amassing in volume, both identifying sources and processing information become increasingly challenging tasks~\cite{liu2012sentiment}. Sentiment analysis concerns itself with (computationally) extracting subjective information regarding entities such as products, locations, or people and encapsulates the analysis of opinions, sentiments, evaluations, appraisals and attitudes~\cite{liu2012sentiment}. From rule-based to statistical approaches and recent developments such as deep learning, research in this space is flourishing, driven by commercial incentives and the far-reaching domains in which it can be applied. Due to the varying sentiment analysis methods between domains, a single cross-domain state-of-the-art accuracy to aspire to does not exist. Furthermore, the subjective nature of opinions as it surfaces in inter-annotator agreements should be considered. If we for example consider an 80\% annotator agreement, it implies that humans will disagree with \textit{any} assigned label about 20\% of the time. Finally, it should be noted that the goal of this research is not to build the perfect sentiment classifier for the Bitcoin domain, but rather select classifiers of satisfactory accuracy (ideally within a 10\% range of the benchmarks) for further analysis of sentiment in relation to market movements.  
 
Early work on machine learning based sentiment analysis methods was carried out by Pang and Lee in~\cite{pang2002thumbs}, wherein it was shown that machine learning techniques such as Naive Bayes, Logistic Regression, and Support Vector Machines on unigram and bigram features could rival human-generated baselines on movie reviews (82\% accuracy). By analyzing the data with these algorithms first, we can discover whether there is a correlation and causal relation at all. If there is, future research could investigate whether improving the sentiment classification can strengthen this correlation, such that misclassification does not propagate to trading models.

\subsection{Data Labeling}\label{sec:data-labeling}
Labeling representative data is an integral part to natural language processing and machine learning processes which try to learn from a correctly labeled dataset. In supervised machine learning, models are first trained on data containing the correct labels so that the models can learn a classification function from the features. As we try to determine whether news articles or social posts regarding Bitcoin are positive or negative, we first show the classifiers what features belong to positive and negative classes through training sets. Based on the labels and corresponding features the model has seen during the training process, it tries to estimate a label for a feature set belonging to an unlabeled item. Sometimes labels can be inferred from context or meta-data. In sentiment analysis of product reviews, the star rating might be used as a proxy label for the review text. Unfortunately the collected data is not augmented with labels or relevant meta-data and needs to be annotated. 

Linguistic annotation tends to be carried out by experts and is expensive and time consuming. It then seems tempting to resort to cheaper and faster methods. However an evaluation of these methods should be taken into consideration.~\cite{Snow2008} compares expert annotations to annotations provided by Amazon Mechanical Turk (http://mturk.com) workers for affective text analysis on headlines of news articles. Amazon Mechanical Turk (hereafter MTurk) is an online platform where requesters can employ workers on human intelligence tasks (HITs), ranging from classification and categorization to collecting feedback and moderating content~\cite{mturkfaw}. The platform includes a dedicated service for sentiment annotation. Requesters can upload a dataset they want to annotate and specify rating scales and instructions.~\cite{Snow2008} found that experts agree with each other more than non-experts agree with experts. Individual experts were found to provide better annotations than non-expert annotators. However, on average 4 non-expert annotators can rival the annotation accuracy of a single expert on affective text analysis tasks. Similarly, in~\cite{mellebeek2010opinion} it is reported that for sentiment analysis, three MTurk annotations per item  provide similar or better performance compared to a single expert annotation. These insights were used as a starting point for the MTurk process in this paper in that more than 4 non-expert annotators were assigned to each document. However, increasing the number of annotators introduces disagreement, leaving us with the challenge of inferring the true label, given a set of assigned labels by various annotators. In doing so, annotator bias as well as ambiguity in the task have to be taken into account. Furthermore, there does not have be a single true label for each item, as some are more open to interpretation than others. 

Leaving out low performing workers should increase the overall precision of assigned labels, especially when these low-performing workers have provided a large number of annotations. The motive for low quality submissions should be clear; workers are paid based on the number of tasks completed, and with no gold standard to which their submissions can be compared from the start, workers have to make a trade-off between speed and quality to optimize their earning. A simple frequentist approach may suggest the use of a majority vote, but this does not take individual noise into account.~\cite{dawid1979maximum} discusses the inference of a true label by means of expectation maximization. In~\cite{Whitehill2009} this approach is applied specifically to Amazon Mechanical Turk by  modeling the ability of individual workers. 

\begin{figure}
  \centering
  \includegraphics[width=0.8\columnwidth]{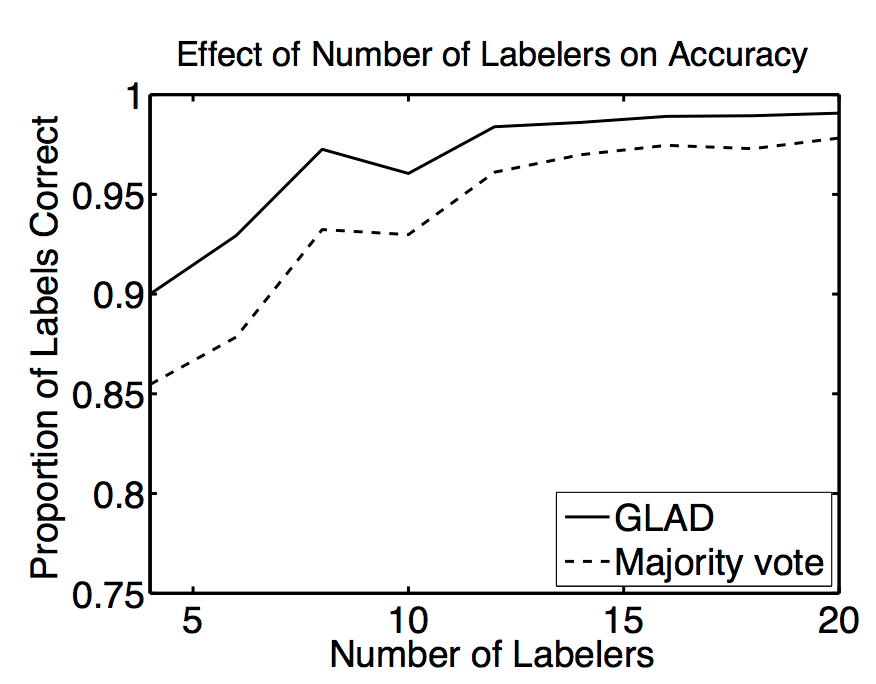}
  \caption{Majority vote versus GLAD approach as proposed by~\cite{Whitehill2009}.}
 \label{fig:glad}
\end{figure}

As figure~\ref{fig:glad} demonstrates, this approach will achieve a maximum improvement of 5\%. While this could be significant when the labels used in this research were to be applied to practical trading algorithms, for now it was decided to only determine whether there is a correlation between sentiment and exchange first, and then look at possible methods of strengthening this correlation if worthwhile. 

\section{Approach}\label{sec:approach}
\begin{figure*}
  \centering
  \includegraphics[width=\textwidth]{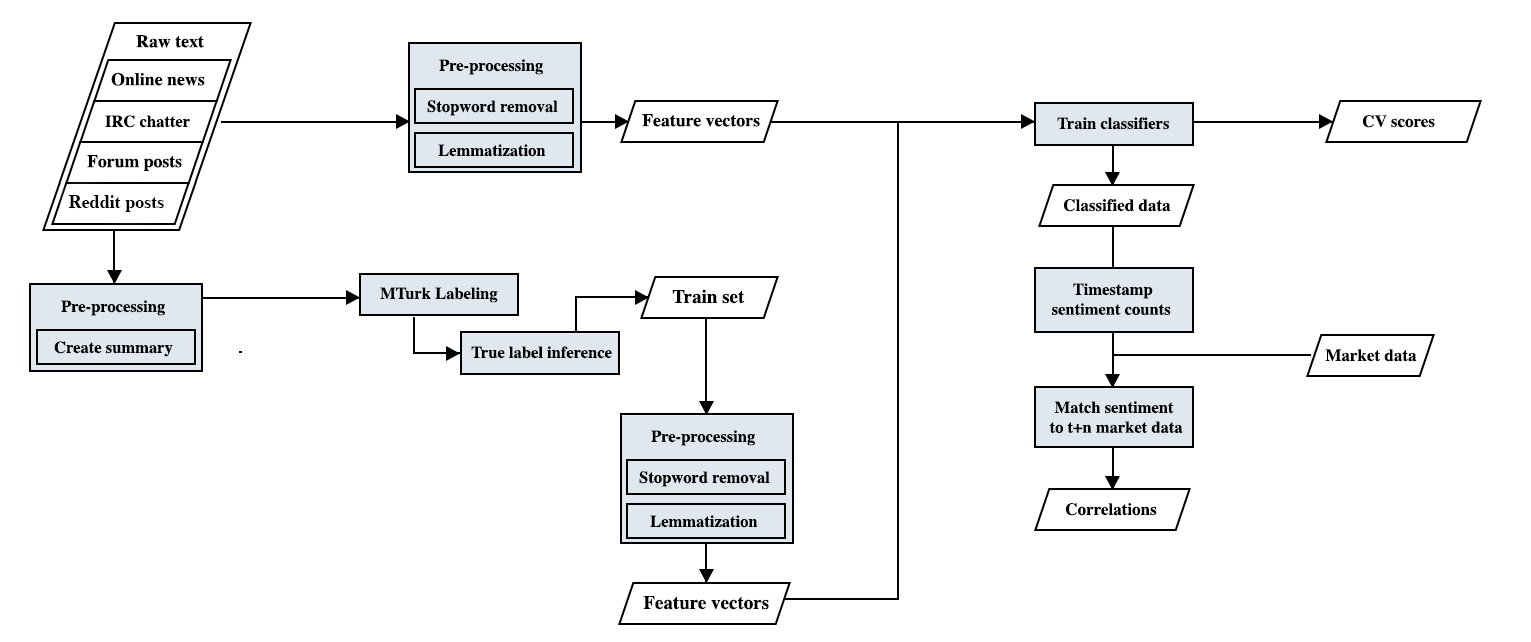}
  \caption{Flow diagram of the NLP process}
  \label{fig:flowdiagram}
\end{figure*}

The following sections will describe the data analysis process from data collection to classification in detail. First, the collection of data from online channels is described. Once data has been collected, a randomly selected subset of the data for each channel will get labeled by gathering crowd-sourced labels for each item in the selected subset. By collecting multiple labels for each item, we will be able to increase the likelihood of the true label being inferred. This labeled set will be preprocessed before transforming it into numerical feature vectors. Likewise, transformations are applied to the unlabeled dataset. Consecutively, the classifiers used to learn from the labeled data will be described. The performance of each individual classifier is then compared. As some classifiers are expected to perform better on data from a particular source compared to others, the classifier with the highest cross-validation per source will be selected to classify the unlabeled data for the respective channel. The output of this classification will be matched on daily timestamps to market data to determine whether there is a correlation between the amount of positive and negative online discourse and the upward and downward exchange movements on \textit{n} days after the recorded sentiment occurred. Figure~\ref{fig:flowdiagram} illustrates the approach in detail.

\subsection{Collected Data}\label{sec:collected-data}
\subsubsection{News Reports}\label{sec:collected-news}
News articles were collected from Bloomberg, Reuters, Coindesk, news.bitcoin.com, Wallstreet Journal, and CNBC, yielding a total of 7,730 articles, of which 1,534 were published in 2015. Besides the article body, the author and the date on which the article was published have also been collected. To select only Bitcoin-related articles, the custom-built web scraper would either use the corresponding website's search function and traverse through all pages, or filter on article tags where available. The scraper did not discriminate on publication date and collected articles that were tagged with Bitcoin or showed up when searching for Bitcoin using the website's search function. Scraping articles from a wider date range (2012-2016) will allow us to train the classifier on a wider breadth of vocabulary; it could be possible that Bitcoin was plagued many highly similar negative events in a specific period, and that the vocabulary in describing those events largely centers around a single topic with the same sentiment. Training a classifier on such sets would limit the possible application of this research in trading models as it generalizes poorly to the wider domain and will therefore not be reliable in assessing unfamiliar events.  

\subsubsection{Forum and Reddit posts}\label{sec:collected-reddit}
With more than 550,000 topics and 1,500,000 posts since its inception in 2009, Bitcointalk.org is by far the biggest Bitcoin-related forum. Irrelevant subforums where off-topic discussions take place were filtered out by excluding URLs from those subforums from being scraped, before mining posts from the following subforums; `Speculation', `Economics', `Trading discussion'. These three subforums contain discussions that are directly related to trading in Bitcoin, and have active market participants posting topics and replies. For forum topics, the topic title and body, the timestamp as well as the author name and total number of replies were indexed. The same data was collected from Reddit, but also included the score of the Reddit post (a function of `upvotes' and `downvotes' by users in the community). 

\subsubsection{IRC Chat}\label{sec:collected-irc}
The two biggest IRC (Internet Relay Chat) channels were scraped for text messages through the BitcoinStats.com website over 2015. For each message, the author, content and timestamp was recorded. Due to the informal writing style that is prevalent in online chat rooms, the data collected from the Bitcointalk forum and Reddit, as well as IRC will likely be significantly noisier than that collected from news articles.

\subsubsection{Market Data}\label{sec:collected-market}
Market data was collected through Blockchain.info through the CSV download option~\cite{bchaininfo}. Blockchain.info is the sole source upon which this paper relies for data relevant to the fundamentals of Bitcoin being traded on exchanges, and contains the daily data from the Bitfinex, Bitstamp and BTC-e exchanges.  

The downloaded CSV  contains the following data, per day over 2015:
\begin{itemize}
\item  Date (daily timestamp)
\item  Average (average  price on daily timestamp)
\item  Ask  (average ask price, sell offer, on timestamp day)
\item  Bid  (average bid price, sell offer, on timestamp day)
\item Last  (Last price recorded on day of timestamp)
\end{itemize}

\subsection{Crowdsourced data labeling}\label{sec:crowdsourced-labels}
In order to quickly label the news articles, the Amazon Mechanical Turk sentiment analysis service was employed to create a labeled corpus of 1,000 randomly selected news articles.  Using the Amazon Mechanical Turk service, English-speaking workers were asked to rate provided sentiment. For our task, a 5-point scale ranging from `very negative' through `neutral' to `very positive' was used. By including a task description and title, workers could opt out of this task for any reason before starting. 

In general, sentiment analysis aims to determine the attitude or polarity of the content with regard to some topic. The topic for the presented content will be Bitcoin. 
Due to the massive amount of different workers, there is a high likelihood that the vast majority of them do not have any expert knowledge of Bitcoin or trading. Instead of directly asking workers what the impact of the content would be on the exchange price, they were asked to rate the perceived effect on the public opinion towards Bitcoin. By doing so, we aim to discover the true intended emotional communication of the content. Negative news about Bitcoin can inherently not have a positive effect on the public opinion towards it. To aid in rating the sentiment for the presented content, workers were provided examples for each item on the rating scale. A detailed view of this task setup can be found in the appendix. The direct task prompt states that the user should assume that the content is Bitcoin-related. Besides that, the task was kept as succinct as possible. 
\newline
\newline
\textbf{`'Assuming that the content is related to the digital currency Bitcoin, rate what kind of impact will this content have on the public opinion towards Bitcoin (BTC)`'} 
\newline
\newline
The direct prompt that was posed to the human intelligence workers is not a true sentiment analysis prompt, in that it assumes that the content is related to Bitcoin and that it asks for the impact on the public opinion towards Bitcoin. As such, the trained classifiers will not classify sentiment of the text, but rather try to classify the perceived intended effect of the article on the Bitcoin public opinion. This is valuable when applying this information in trading strategies, as traders will try to judge what decision other players in the market will make upon the release of the relevant information. Articles in which a Bitcoin competitor is mentioned negatively would have a negative classification in traditional sentiment analysis tasks, but might be positively classified by the human intelligence workers responding to the prompt. 

\begin{figure}
  \centering
  \includegraphics[width=\columnwidth]{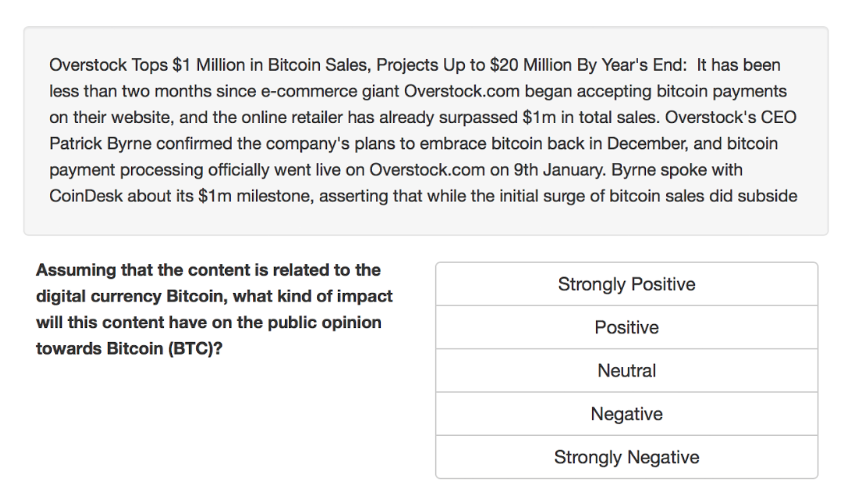}

  \caption{Screen capture of an example task as it is presented to MTurk workers. }
  \label{fig:screencapture}
\end{figure}

Content for the HIT tasks was created by taking the article headline or topic title and the first 500 characters of the article or topic body where available. Titles of financial articles tend to give a summary of the general expressed sentiment in the article. Some forum topics consist solely of an expressive title, and in such cases only the title was used. The first 500 characters of the article or post body are included to provide context on the title. This 500-character limit was imposed to not overload MTurk workers with massive walls of text; analyzing this text would require more time as confounding information might be given in the full article, consequently increasing the costs of the labeling task.  Figure~\ref{fig:screencapture} shows an example task as presented to HIT workers. 

As the syntactical and semantic structure of generally well-composed news articles differs greatly from the structures used on online forums and chat channels, this annotation process was performed on 2,000 forum topics, 2,000 Reddit posts, and 2,500 IRC messages. Sentiment classifiers on each source will be trained separately.  

The task setup was the same for news articles, forum topics, and Reddit posts. The approach for IRC chatter was different, as messages are typically very short and lack context compared to the former. To adjust for this, workers were simply asked to rate whether the general content of the message had an expressed sentiment ranging from very negative through neutral to very positive, disregarding the Bitcoin context. 

\subsection{Mechanical Turk result analysis}\label{sec:mturk-results}
\begin{figure}
  \centering

  \includegraphics[width=\columnwidth]{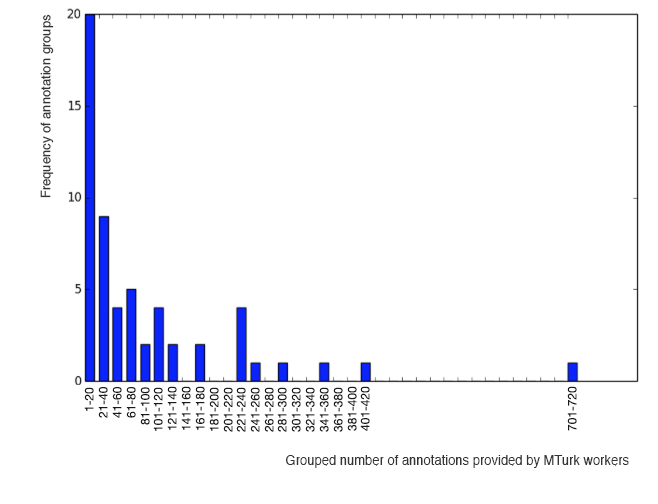}
  \caption{Worker activity distribution on news articles. Annotations per worker are grouped into bins of 20. Bins without any annotations in the 421 - 701 range are unlabeled.}
  \label{fig:mtv}
\end{figure}

The labeling of each dataset was completed within 48 hours of task submission. Workers were compensated with \$0.02 for each provided label. An analysis of the labeled news article data shows that 57 workers provided a total of 5000 annotations (each item was labeled 5 times), for an average of 87.7 annotations per worker. The standard deviation of this population is 121.67, indicating a high variance in the amount of provided annotations between workers. Due to this high variance, it is important to control for individual worker bias. The graph of this variance is shown in Figure~\ref{fig:mtv}.

Table~\ref{tab:mturk-analysis} shows a distribution of the labels provided by workers, aggregated over all 5,000 annotations provided. Very negative, negative, neutral, positive, and very positive annotations are indicated by the column headers - -, -, 0, +, + +, respectively.  With 37.62\% of instances being assigned 'neutral', and a slight bias towards positive news indicating a balanced news landscape. While all news outlets report more positive than negative news, WSJ, Bloomberg, news.bitcoin.com and Bitcoin report about three or more times as many positive than negative reports. This could potentially indicate a bias in their reporting, but can also be caused by the sampled content; the value of Bitcoin has steadily increased over the years the sampled content was published in.  This bias is less pronounced in the forum, and IRC annotations.  It can further be noted that the largest amount of labels for each source were assigned a neutral label. 

\begin{table}
  \centering
  \caption{Annotation distribution across news channels.}\label{tab:mturk-analysis}
  \scalebox{0.8}{
  \begin{tabular}{cp{10mm}p{10mm}p{10mm}p{10mm}p{10mm}}\toprule
             &\textbf{- -} &\textbf{-} &\textbf{0} &\textbf{+} &\textbf{+ +}\\\midrule
    Coindesk & 90 &278&750&852&55\\
    Reuters & 168 &494&868&589&31\\
    NewsBitcoin & 7&43&139&147&9\\
    Bloomberg & 12&45&127&99&2\\
    WSJ & 4 &15 &47&56&3\\
	CNBC & 20&83&134&148&25\\
    Total &276 (5.5\%)&847 (16.9\%)&1,881 (37.7\%)&1,876 (37.5\%)&120 (2.4\%)\\
    \bottomrule
  \end{tabular}}
\end{table}

\begin{table}
  \centering
  \caption{Annotation distribution across Bitcointalk subforums}\label{tab:mturk-forum}
  \scalebox{0.8}{
  \begin{tabular}{cp{10mm}p{10mm}p{10mm}p{10mm}p{10mm}}\toprule
             &\textbf{- -} &\textbf{-} &\textbf{0} &\textbf{+} &\textbf{+ +}\\\midrule
    Speculation & 147 &1,087&2,635&1,017&124\\
    Economics & 75 &707&2,925&1,163&120\\
    Trading & 51&417&3,260&1,250&21\\
    Total &273 (1.8\%)&2,211 (14.7\%)&8,838 (58.9\%)&3,430 (22.8\%)&265 (1.8\%)\\
    \bottomrule
  \end{tabular}}
\end{table}

\begin{table}
  \centering
  \caption{Annotation distribution across Reddit pages}\label{tab:mturk-reddit}
  \scalebox{0.8}{
  \begin{tabular}{cp{10mm}p{10mm}p{10mm}p{10mm}p{10mm}}\toprule
             &\textbf{- -} &\textbf{-} &\textbf{0} &\textbf{+} &\textbf{+ +}\\\midrule
    bitcoin & 66&535&3,206&1,074&96\\
    btc & 101 &600&3,051&1,150&111\\
    Total &167 (1.7\%)&1,135 (11.4\%)&6,257 (62.6\%)&2,224 (22.2\%)&217 (2.2\%)\\
    \bottomrule
  \end{tabular}}
\end{table}

\begin{table}
  \centering
  \caption{Annotation distribution across IRC channels}\label{tab:mturk-irc}
  \scalebox{0.8}{
  \begin{tabular}{cp{10mm}p{10mm}p{10mm}p{10mm}p{10mm}}\toprule
             &\textbf{- -} &\textbf{-} &\textbf{0} &\textbf{+} &\textbf{+ +}\\\midrule
    otc & 133&1,590&8,944&1,634&54\\
    dev & 71 &1,877&8,163&2,260&59\\
    Total &204 (0.8\%)&3,467 (14\%)&17,107 (69\%)&3,894 (15.7\%)&113 (0.5\%)\\
    \bottomrule
  \end{tabular}}
\end{table}
Applying a simple majority vote with the aim of inferring the true labels for a given news article leads to a more pronounced bias towards positive news, outweighing negative news by a factor of 3 to 1, and increasing the relative size of the neutral class. The results can be seen in Table~\ref{tab:majorityvote}. 

\begin{table}
  \centering
  \caption{Majority vote label distribution}\label{tab:majorityvote}
  \begin{tabular}{cp{10mm}p{10mm}p{10mm}p{10mm}}\toprule
\textbf{- -} &\textbf{-} &\textbf{0} &\textbf{+} &\textbf{+ +} \\\midrule
16    (1.6\%)&122 (12.2\%)&479 (47.9\%)&379 (37.9\%)&4 (0.4\%)\\
    \bottomrule
  \end{tabular}
\end{table}

The majority vote on news data causes 47.9\% of samples to fall into the neutral and furthers the imbalance between positive and negative classes. We can infer that such a majority vote on the other datasets will have a similar if not more pronounced effect due to the higher presence of neutral labels. The true difference will be established in section~\ref{sec:sentiment-classification}, where the errors of each classifier are analyzed. As the prediction we make is binary (i.e. positive or negative), it is expected that leaving out the neutral label will lead to a clearer decision boundary between nearby labels in binary prediction. Before continuing, 'very positive' and 'positive' labels were merged into 'positive'.

\subsection{Feature Extraction}\label{sec:feature-extraction}
What follows is an overview explaining the applied preprocessing and feature extraction techniques. 

\subsubsection{Pre-processing Data}\label{sec:preprocessing} 

\subsubsection{Lemmatization}\label{sec:lemmatisation}
Lemmatization is a normalization process by which morphological variation in data can be reduced. Word forms such as `diving' and `dove' will be mapped to their dictionary form `dive'. Compared to the more crude process of stemming (which is faster and less complex) the produced lemmas are linguistically valid~\cite{manning2008introduction}. 
		 				
To create accurate lemmas, each token has to be assigned a part-of-speech tag (POS tag) before determining the lemma. In POS-tagging lexical categories such as `Noun' or `Adverb' are added to each word. These lexical categories are valuable in handling disambiguation. The spaCy Python package~\cite{spacyweb} has been used to tokenize, parse, and lemmatize the data from each source. The Spacy parsing accuracy is within 2\% of the state-of-the-art parser by introduced by~\cite{andor} (92.8\% for Spacy, 94.3\% for Andor et al.) according to a comparative study carried out by Choi et al.~\cite{choi2015depends}. Settings were left on default. 

Texts may also contain various types of noise such as stop words and punctuation that may not have any influence on the polarity of the text. Before processing the data with classifiers, the data was cleaned. In the process of filtering stop words, negating terms were removed from the NLTK stop word corpus~\cite{stopwordcorpus}. Punctuation was removed using a regular expression, regardless of the source channel. However,~\cite{saif2014stopwords} postulates that removal of stop words from tweets has a negative influence on sentiment prediction accuracy due to the noisy nature of short text messages (abbreviations and irregular forms). In the collected datasets for this research, there was no demonstrated improvement upon stop word removal, and they were left in place.

The collected text documents have to be converted to numerical feature representations so that they can be used by the statistical classifiers. Text is converted to word n-grams (features), and then represented using a bag-of-n-grams approach. 

Using the Scikit learn~\cite{scikitlearnweb} CountVectorizer feature with n-gram range specifications, a vectorizer for unigrams, bigrams and trigrams was created. The corpus is then transformed to a sparse matrix counting the occurrences of each word n-gram per document, using the respective vectorizer for each of the n-grams. 

The theoretical benefit of bigrams and trigrams over unigrams, is that they maintain word order. In the context of sentiment analysis the unigram features of the sentence 'I do not like Bitcoin' would be ['i','do', 'not', 'like', 'bitcoin']. A trigram would contain the feature [..., ['not','like','bitcoin'], ...], and would include the connection that it is the Bitcoin that is being disliked. A bigram containing [‘not’, ‘like’] allows for negation. This context is lost on unigrams and BOW approaches. 

\subsubsection{TF-IDF Transformation}\label{sec:tfidf}
To counter the overly present words in this specific domain, the counts in the feature vectors are weighted and selected using a TF-IDF transform~\cite{salton1988term}. This is different from stop word removal, as it is adaptive to domain contexts.  If the resulting matrix from vectorization would be directly used in classification, common words would outweigh the less common but more relevant words. The main idea of TF-IDF is that a word that is very common in a specific article, but does rarely appear in other articles, will provide a strong indication that this word will offer good categorical differentiability. 

\begin{equation}
w_{i,j} = tf_{i,j} \times  \log(\frac{N}{df_{i}})
\end{equation}

This process will return a weight \(w_{i,j}\) multiplying the term frequency \(tf\), the number of times a term \textit{i} occurs in document \textit{j}, by the inverse document frequency (the total number of documents (\textit{N}), divided by the total number of documents containing term i (\(df\)). 

The performance and added value of the TF-IDF transform has been evaluated against the collected content, and has not shown an increased cross-validation performance for any discourse channel combined with any of the proposed classifiers. A possible explanation for this, is the relatively short texts which are being classified (recall that they were limited to the title plus the first 500 characters). TF-IDF performs well on longer documents, but short text is known to cause noisy TF-IDF values~\cite{jain2014short}.

\subsection{Sentiment classification}\label{sec:sentiment-classification}
The application of each of the three classifiers (AlchemyAPI, logistic regression, Naive Bayes) will be explained with the news data as an illustrative example. Summarized results for all sources are presented at the end of this section, as the approach is the same for each source. Full results for each source can be consulted in the appendix.

\subsubsection{AlchemyAPI}\label{sec:alchemy-api}
We start by establishing a baseline for our sentiment classification using the AlchemyAPI (part of the IBM Watson cloud offering), before employing classifiers from the scikit library. AlchemyAPI is a natural language machine learning service~\cite{alchemy}. The Alchemy Language API provides out-of-the-box solutions for sentiment analysis, but can not be trained to suit a particular domain and might suffer from the domain-transfer problem~\cite{domaintranfer}. 

The raw text data was sent to be processed by the AlchemyAPI the same way it was sent to the MTurk annotation service, as the AlchemyAPI has its own text pre-processing in place. This means that the pre-processing as applied in the previous section is irrelevant to this classifier. It should be noted that the way in which the data was labeled might affect the evaluation, as AlchemyAPI is a pure sentiment analysis API, whereas annotators were asked to label items according to the perceived effect on the public opinion towards Bitcoin. 

\subsubsection{AlchemyAPI Error Analysis}\label{sec:alchemy-api-errors}
From the confusion matrix in Table~\ref{tab:alchemytftable}, we can determine that the AlchemyAPI achieves a recall of 74.2\%, and a precision of 81.3\% on summarized news articles, using the average of MTurk annotations as gold standard. 
Recall is defined  to be the ratio of correct classifications divided by the total number of correct classifications.  
\begin{equation}
\label{eq:recall}
Recall =  \frac{True Positives}{True Positives + False Negatives}
\end{equation}
In the context of trading applications and a hypothetical integration into trading models, it is absolutely critical to maximize precision, as uninformed trades could lead to significant financial losses. It does not matter that some profitable trades are being missed (recall), we just want to ensure that the trades that are made are indeed correct. This is not to say that recall does not matter at all, one would still want to make enough profitable trades to outperform the benchmark of simply buying and holding Bitcoin.

\begin{equation}
Precision =  \frac{True Positives}{True Positives + False Positives}
\end{equation}

When analyzing misclassification (17\% of true positive instances, 34.89\% of true negative instances), the classification of negative instances performs significantly less than that of positive instances. In accounting for this, it is important to keep the bias of individual news sources in mind. A further possible explanation for this may be the wide ranging interpretations that can be given to a Bitcoin-related text. For example, a document could receive the labels [-2, -1, 0, 0, 1] from five individual annotators and subsequently receive a neutral label due to the majority vote. However, just as many annotators gave it a neutral label as a negative label, the negative annotators simply did not agree on the degree of negativity. The negative sentiment assignments are then completely disregarded in the final label inference. This negative sentiment is however still contained in an average of the five labels.    

\begin{table}[]
\centering
\caption{AlchemyAPI Confusion Matrix}
\label{tab:alchemytftable}
\begin{tabular}{@{}lll@{}}
\toprule
\textsubscript{Pred.} / \textsuperscript{Real} & \textbf{P} & \textbf{N} \\ \midrule
\textbf{P}           & 422  & 127  \\
\textbf{N }          & 87  & 237  \\ \bottomrule
\end{tabular}
\end{table}

\subsubsection{Logistic Regression}\label{sec:logisticregression}
Logistic regression (also called maximum entropy or MaxEnt) has proven effective in various text classification tasks~\cite{berger96} and is often seen as the go-to method for binary classification problems. It is easy to implement and does not require any tuning. For now, we will again resort to the implementation as it is provided in the scikit-learn Python library. At its core, logistic regression is based on the logistic function which takes any input \(x\) and maps it to a value between the limits 0 and 1.

\begin{equation}
\frac{1}{1+e^{-z}}
\end{equation}

As the predictions are boolean, the logistic regression model for determining the probability of feature set \textit{X} containing individual features \(x_{i}\) belonging to class \textit{1}, \(P(1|X)\), can be given as

\begin{equation}
P(1|X) = \frac{1}{1 + e^{w_{0} + \sum\limits_{i=1}^n w_{i}X_{i} }}
\end{equation}

with \(x_{i}\) representing each feature in the feature space. The weights \(w_{i}\) of each feature \(x_{i}\) are determined using a Maximum Likelihood Estimator.

As the probabilities must sum to 1, \textit{P(0|X)} is given by 

\begin{equation}
P(0|X) = \frac{e^{w_{0} + \sum\limits_{i=1}^n w_{i}X_{i} }}{1 + e^{w_{0} + \sum\limits_{i=1}^n w_{i}X_{i} }}
\end{equation}

We can now assign label \textit{0} if the  below condition holds. Otherwise label \textit{1} is assigned.
\begin{equation}
1 < \frac{P(0|X)}{P(1|X)} 
\end{equation}

A first pass with logistic regression on word unigram features shows a 78\% and 73\% and 71\% accuracy rating for news articles, forum topics and Reddit posts respectively. Unigrams achieving the highest accuracy is in line with the findings of Pang and Lee~\cite{pang2002thumbs}. The under-performance of n-grams compared to unigrams can be attributed to the limited availability of labeled data (which will then be further split into train and test sets in cross-validation). Unique bi- and tri-grams are anticipated to occur less often across a sparse set of documents. This causes the classifier to come across more 'unseen' n-grams than unigrams.

\subsubsection{Logistic Regression Error Analysis}\label{sec:logregerrors}
Inspecting the classification errors shows that the majority (66.67\%) of misclassified news documents contain `neutral' as the majority vote assigned by MTurk workers. 
When comparing majority votes and assigned average scores and removing neutral labeled documents from the corpus, we find that an average of all five labels yields higher cross-validation scores, because this effectively leaves out the neutral label in a large number of instances. A mere 16 instances received all neutral annotations, meaning that the rest would, on average, have either a positive or negative sentiment. The majority vote greatly increases the number of neutral labels, up to a third of the total dataset. The classifier output is binary for positive or negative, and the third neutral class will never be predicted. Multi-class classification generally also suffers from higher error rates and in~\cite{zhang2015text} it is also argued that maintaining a third neutral label would blur the decision boundary between positive and negative and decrease performance. For the above reasons, the average of MTurk annotations is used as the correct label in the classifier training process.  This effectively eliminates neutral instances (very few documents received all-negative labels by all five annotators).  Fully neutral documents were removed from the training set. 

\subsubsection{Naive Bayes}\label{sec:nb-classifier}
Naive Bayes (hereafter NB) is often considered the baseline for many text classification models as it is robust, accurate and fast to implement. Traditional NB models assume that all attributes of the sample are independent to each other. Although often false (relations between words offer context, which is lost in a simple bag-of-words approach), NB performs well in various real-world tasks. Domingos and Pazzani explain this apparent paradox in~\cite{domingos1997optimality} by asserting that classification estimation is only a function of the sign (1, 0) of the function estimation. NB relies on Bayes' theorem to compute the posterior probability of a class, given the distribution of features in the input vector.  

\begin{equation}
\label{eq:pfrac}
P (C_{i} | F_{j}) = \frac{P(F_{j} |  C_{i}) P(C_{i})}{P(F_{j})}
\end{equation}

In equation~\ref{eq:pfrac}, \(P(C_{i})\) is the prior probability of class \textit{i} existing, independent from any other factors. \(P(C_{i}|F_{j})\) is the prior probability that a given feature set \(j\) is classified as \(C_{i}\).\( P(F_{j})\) is the prior probability that a given feature set occurs, again independent from any other factors.

Two common implementations of the NB model for text classification are the Bernoulli and Multinomial models. They differ in the method in which the features they use are represented. In the Bernoulli model, each document is represented by a binary vector over the word space, where dimensions match words from the vocabulary, indicating only whether or not a word is present in the document. Multinomial models on the other hand will also consider how often each word appears. The feature vectors built in the previous step do contain word counts and a multinomial model therefore seems better suited. Multinomial Naive Bayes computes the probability of a document \textit{d} belonging to class \textit{c} as follows~\cite{manning2008introduction}:

\begin{equation}
(c|d)\propto log P(c) + \prod_{1\leq k\leq n_{d}} log P(t_{k}|c)
\end{equation}

\(P(t_{k}|c)\) is the conditional probability of feature \(t_{k}\) being present in \textit{c}, and signifies the amount of evidence feature \(t_{k}\) contributes in determining whether this document belongs to class \(c\). \(P(c)\) is the prior probability of any document at all belonging to class \(c\).  The features in the vector of document \(d\) are represented by \(t_{k}\), where \(n_{d}\) is the total number of features in \(d\). The multinomial NB classifier aims to maximize P based on the data that was used to train the model, effectively aiming to select the most likely class from the set of classes, given a certain set of features for the respective document. According to~\cite{manning2008introduction}  multiplying probabilities can lead to floating point underflow. For this reason, the log probabilities are added instead of  multiplied. 

\begin{equation}
\label{eq:argma}
arg max_{c\in \mathbb{C} } \hat{P} (c|d) = arg max_{c\in \mathbb{C} } log\hat{P}(c) + \prod_{1\leq  k \leq n_{d}}  log\hat{P}(t_{k}|c)
\end{equation}

In equation ~\ref{eq:argma} \(\hat{P}\) is used in the above notation as the probabilities are not truly known, they are based on observations made from the training set. 

A comparison confirms the assumption of multinomial Naive Bayes outperforming Bernoulli Naive Bayes. Cross-validation scores showed a 4\% increase Bernoulli to Multinomial using unigram features on the created news article summaries in table~\ref{tab:nbcomp}. This corresponds with the findings in~\cite{wang2012baselines}.

\begin{table}
  \centering
  \caption{Comparison of multinomial NB and Bernoulli NB classifiers on news article summaries }\label{tab:nbcomp}
  \begin{tabular}{cp{25mm}p{25mm}p{25mm}p{25mm}}\toprule
     &\textbf{Multinomial NB} &\textbf{Bernoulli NB} \\\midrule
CV-score    &0.82 &0.78 \\
    \bottomrule
  \end{tabular}
\end{table}

Although both logistic regression and Naive Bayes are used for classification, NB is a generative model, whereas logistic regression is a discriminative model. Generative models try to model the underlying probability distribution whereas discriminative models aim to learn the boundaries between classes~\cite{ng2002discriminative}. Logistic regression splits feature space linearly, and performs well even if some of the features are correlated (NB assumes independence).~\cite{vapnik1998statistical} states that it is preferable to select a discriminative model where possible, as one should aim to solve the classification problem directly, rather than addressing a more general problem (modeling the underlying distribution) as an intermediate step in the classification. 

\subsubsection{Multinomial Naive Bayes Error Analysis}\label{sec:mnberrors}
To maintain a consistent comparison, the average of MTurk annotations is again used as the target in classifier training. Assessing the performance of the multinomial Naive Bayes classifier from the confusion matrices, we find that this classifier is better at predicting negative news than positive news, which seems counter-intuitive upon inspecting table~\ref{tab:labeldis}, given that there are more true positives in the training set. The same holds for all other sentiment channels (table~\ref{tab:mnbredditcm},~\ref{tab:mnbforumcm},~\ref{tab:mnbirccm}).
Table~\ref{tab:labeldis} shows the distribution of target labels for each channel. Class imbalance has decreased compared to the majority vote in table ~\ref{tab:majorityvote}. This inconsistency indicates that the negative class has stronger predictive features than positive class; given the same document length (title and first 500 characters), the features present in a negative document vector provide more evidence of the document belonging to the negative class than the features of a positive document. 

Manual inspection of the errors does not show a clear pattern in document topic of misclassified instances.

\begin{table}
  \centering
  \caption{True class distributions for each sentiment channel }\label{tab:labeldis}
  \begin{tabular}{cp{11mm}p{11mm}p{11mm}p{11mm}}\toprule
     &\textbf{News} &\textbf{Forum} &\textbf{Reddit}&\textbf{IRC}\\\midrule
Positive    &515 (62.80\%) &1.257 (64.42\%) &979 (66.68\%) &1.956 (56.66\% \\
Negative    &305 (37.20\%) &694 (35.58\%) &489 (33.32\%) &1.496 (43.34\%) \\
    \bottomrule
  \end{tabular}
\end{table} 

\subsubsection{Classifier Comparison}\label{sec:clfcomparison}
Table~\ref{tab:newsclf} compares precision, recall, F-measure, and cross-validation score (accuracy) between the applied classifiers on news data with unigram feature vectors. The same is repeated for forum posts (table~\ref{tab:forumclf}), Reddit posts (table~\ref{tab:redditclf}), and IRC chatter (table~\ref{tab:ircotcclf} and ~\ref{tab:ircdevclf}). The process was further split by separating the IRC channels with the assumption that vocabulary might differ between the development (\#dev) and over-the-counter (\#otc) channel. The number of cross-validation folds (10) was chosen iteratively for the logistic regression and multinomial Naive Bayes classifiers, optimizing for accuracy.  A confusion matrix of each source and classifier can be found in the appendix. 

Although the AlchemyAPI is easily scalable, does not require any preprocessing, and shows a decent accuracy score on IRC chatter and forum posts compared to the other models, the low precision and recall make it a poor choice for our application. 

Without a true cost attached to mistaken decisions, a trade-off between precision and recall can not directly be made. The harmonic mean of precision and recall (the F-Measure, introduced by van Rijsbergen~\cite{van1979information}) and accuracy will serve as the main selection criteria for now. The harmonic mean of the precision and recall is calculated as follows: 

\begin{equation}
F1 = 2 \cdot \frac{precision \cdot recall}{precision + recall}
\end{equation}
\begin{table}[]
\centering
\caption{News classifier comparison.}
\label{tab:newsclf}
\scalebox{0.7}{
\begin{tabular}{lllll}
\rowcolor[HTML]{C0C0C0} 
News                                                            & \textbf{Precision} & \textbf{Recall} & \textbf{F-Measure} & \textbf{CV / Accuracy} \\ \hline
\begin{tabular}[c]{@{}l@{}}AlchemyAPI\\ Raw text\end{tabular}   & 0.8290             & 0.7686          & 0.7976             &    0.76                          \\
\rowcolor[HTML]{EFEFEF} 
\begin{tabular}[c]{@{}l@{}}NB MultiNom.\\ unigram\end{tabular}  & 0.782             & 0.728          &0.754                    & 0.82                         \\
\begin{tabular}[c]{@{}l@{}}LogRegression\\ unigram\end{tabular} & 0.775             & 0.626          &0.685                    & 0.78       \\     
\end{tabular}}
\end{table}

\begin{table}[]
\centering
\caption{Forum classifier comparison.}
\label{tab:forumclf}
\scalebox{0.7}{
\begin{tabular}{lllll}
\rowcolor[HTML]{C0C0C0} 
Forum                                                           & \textbf{Precision} & \textbf{Recall} & \textbf{F-Measure} & \textbf{CV / Accuracy} \\ \hline
\begin{tabular}[c]{@{}l@{}}AlchemyAPI\\ Raw text\end{tabular}   & 0.8708             & 0.5685          & 0.7976             &   0.69                           \\
\rowcolor[HTML]{EFEFEF} 
\begin{tabular}[c]{@{}l@{}}NB MultiNom.\\ unigram\end{tabular}  & 0.585              & 0.584           &0.584                    & 0.71                         \\
\begin{tabular}[c]{@{}l@{}}LogRegression\\ unigram\end{tabular} & 0.652              & 0.563           &0.604                    & 0.74    \\                       
\end{tabular}}
\end{table}

\begin{table}[]
\centering
\caption{Reddit classifier comparison.}
\label{tab:redditclf}
\scalebox{0.7}{
\begin{tabular}{lllll}
\rowcolor[HTML]{C0C0C0} 
Reddit                                                          & \textbf{Precision} & \textbf{Recall} & \textbf{F-Measure} & \textbf{CV / Accuracy} \\ \hline
\begin{tabular}[c]{@{}l@{}}AlchemyAPI\\ Raw text\end{tabular}   & 0.8298             & 0.6004          & 0.6967                 & 0.71                          \\
\rowcolor[HTML]{EFEFEF} 
\begin{tabular}[c]{@{}l@{}}NB MultiNom.\\ unigram\end{tabular}  & 0.603              & 0.546            & 0.573                & 0.73                         \\
\begin{tabular}[c]{@{}l@{}}LogRegression\\ unigram\end{tabular} & 0.65              & 0.425           & 0.573                & 0.73       \\
            
\end{tabular}}
\end{table}

\begin{table}[]
\centering
\caption{IRC OTC classifier comparison.}
\label{tab:ircotcclf}
\scalebox{0.7}{
\begin{tabular}{lllll}
\rowcolor[HTML]{C0C0C0} 
IRC \#OTC                                                       & \textbf{Precision} & \textbf{Recall} & \textbf{F-Measure} & \textbf{CV / Accuracy} \\ \hline
\begin{tabular}[c]{@{}l@{}}AlchemyAPI\\ Raw text\end{tabular}   & 0.57677            & 0.6720          & 0.6207                 & 0.77                         \\
\rowcolor[HTML]{EFEFEF} 
\begin{tabular}[c]{@{}l@{}}NB MultiNom.\\ unigram\end{tabular}  & 0.623             & 0.57          & 0.595                & 0.65                         \\
\begin{tabular}[c]{@{}l@{}}LogRegression\\ unigram\end{tabular} & 0.7             & 0.538            & 0.609                & 0.69      \\                     
\end{tabular}}
\end{table}

\begin{table}[]
\centering
\caption{IRC DEV classifier comparison.}
\label{tab:ircdevclf}
\scalebox{0.7}{
\begin{tabular}{lllll}
\rowcolor[HTML]{C0C0C0} 
IRC \#DEV                                                       & \textbf{Precision} & \textbf{Recall} & \textbf{F-Measure} & \textbf{CV / Accuracy} \\ \hline
\begin{tabular}[c]{@{}l@{}}AlchemyAPI\\ Raw text\end{tabular}   & 0.3099             & 0.8549          & 0.45489                 & 0.658                        \\
\rowcolor[HTML]{EFEFEF} 
\begin{tabular}[c]{@{}l@{}}NB MultiNom.\\ unigram\end{tabular}  & 0.641             & 0.59          & 0.616                & 0.69                         \\
\begin{tabular}[c]{@{}l@{}}LogRegression\\ unigram\end{tabular} & 0.623             & 0.631          & 0.628                & 0.71 
\\ 
\end{tabular}}
\end{table}

\section{Correlation and Causation}\label{sec:correlation}
Recall that sentiment was classified in a binary manner. This means that a single negative post cancels out a single positive post when counting sentiment cumulatively, without any regard for the degree of polarity. In order to establish the correlations of cumulative, negative and positive sentiments individually, correlations have been calculated separately. 

Initially, it is assumed that sentiment can correlate with market movements for more than one day after the sentiment is expressed. As we are mostly interested in predicting upward or downwards movement in exchange price and trade volume, we calculate the percentage and absolute changes in volume and average daily price between \(t\) (the date at which the sentiment was expressed) and \(t+n\) (the date on which we calculate the correlation). 

Tables~\ref{tab:poscor},~\ref{tab:negcor}, and~\ref{tab:cmlcor} illustrate correlations of the positive, negative, and cumulative sentiment at \(t\) with the price and volume changes from \(t\) to \(t+n\). For each channel the best respective classifier from the previous section was chosen to classify the dataset over 2015. The Pearson  correlation coefficients \((r)\) and corresponding p-values are calculated using the \textit{pearsonr} module from the Python SciPy library~\cite{scipy}. 

For all channels except Reddit, positive sentiment seems to correlate negatively with changes in trading volume from \textit{t+2} onwards at significance levels of at least 0.05 or below. Interestingly, positive Reddit sentiment is the only sentiment source which positively correlates with upwards price movements. No other positive sentiment channels exhibits such a correlation - all other positive sentiment correlates negatively. The differences in correlations of the same sentiment type between channels could offer interesting insight into how Bitcoin market participants who reside in specific online communities respond to news (which may or may not be exclusive to that community).

Overall, correlations are stronger for negative expressions of sentiment. In the context of the hypothesis, this could indicate that Bitcoin traders are more sensitive to negative news. This matches with the findings of~\cite{tetlock2007giving}. In making that statement, it is important to take the classification errors into account. Incorrect classifications of positive sentiment spill over into the negative class. This means that the classifiers will recognize more negative news than there truly is. The same is true in the other direction, which on average compensates this effect to some degree. Future research improving the classification of positive sentiment will likely decrease the correlations of negative sentiment slightly.  

The strongest correlations can be found in forum sentiment. A possible explanation for this is the subforum filtering that preceded data collection. Other sources discuss anything Bitcoin related, whereas the scraped subforums center around Bitcoin trading, speculation, and economics, and as such could have a larger population of active Bitcoin traders. 

Although the correlations found thus far are weak to moderate, a perfect correlation was never expected; there are many other factors at play in Bitcoin price formation. 

In a first effort of strengthening the correlations, we only consider sentiment at time \textit{t} and market movements at \textit{t+n} if the positive sentiment at \textit{t} is above the 90th percentile of daily positive sentiment for 2015 on that channel and negative sentiment is below the 10th percentile of negative sentiment for 2015. This improved correlations, but decreased the statistical significance (p > 0.05) by large amounts. Due to the small remaining sample (10th percentiles of 365 daily observations over 2015), we can observe correlations even if there is no true correlation between variables at all. The increase of p-values reflects this uncertainty.

\subsection{Granger Causality Test}\label{sec:granger}
To determine whether sentiment has any predictive value in forecasting Bitcoin exchange movements in price and volume, we can perform a Granger causality test~\cite{granger1969investigating}. The Granger causality test determines whether information in one time series (\textit{x}) precedes the other (\textit{y}) and whether including historical data from \textit{x} improves the prediction accuracy of \textit{y} over using only historical data from \textit{y}. It is then said that \textit{x} Granger-causes \textit{y}. The Granger causality test accomplishes this by using the information in one time series to model the changes in the other time series. This test does not aim to prove true or philosophical causality. The test only aims to establish predictive causality~\cite{diebold1998elements}. 

As the tests are bi-directional (we want to determine the direction of causality), the null hypotheses H\textsubscript{0} for the performed Granger causality test are the following:

\begin{itemize}
\item \(H_{0.1}:\)\textit{Series x does not Granger cause series y}
\item \(H_{0.2}:\)\textit{Series y does not Granger cause series x}
\end{itemize}
Series \textit{x} will be the type of sentiment, and series \textit{y} the market metric. Null hypotheses are rejected at significance levels beyond 0.05.

The Granger causality test requires that the analyzed time series are both covariance stationary. There should be no clear (up- or downwards) trends in the data. To determine this, an augmented Dickey-Fuller (ADF)~\cite{mushtaq2011augmented} test is performed on each of the series that will be tested in \(H_{0}\). Stationary series for each variable are then used in the Granger causality test. The ADF test confirms stationarity of percentage daily changes in price and volume, and all collected sentiment time series.

The EViews 9 software~\cite{eviews} was used to perform the causality tests with various lag settings. A lag-length of for example 2 makes the assumption that data from one of the selected time series \textit{x} does not help predict series \textit{y} if it is more than two time steps removed from \textit{y}.

Table~\ref{tab:gcoverview} shows the summarized findings of the Granger Causality tests as performed on news articles, Reddit posts and forum topics, with percentage changes in volume and average daily price. For news and forum channels, we most notably find a Granger-causality from negative sentiment to the price change, indicating that negative news sentiment has value in predicting these movements. This observation is reversed in negative Reddit sentiment, where sentiment is Granger-caused by percentage changes in price. This surfaces an interesting pattern in the behavior of online Bitcoin communities. News and forum are seemingly used to collect trading intelligence, whereas Reddit seems contain discussion as to what happened on the markets. We also find that volume changes are leading indicators for negative and positive forum sentiment and negative Reddit sentiment. None of the  analyzed channels show a predictive causality from negative or positive sentiment to changes in trading volume. 

Full results of the Granger tests can be found in table~\ref{tab:gcnews},~\ref{tab:gcreddit} and ~\ref{tab:gcforum} in the Appendix.   

\begin{table}[]
\centering
\caption{Summarized results of the performed Granger causality tests, showing the direction of Granger-causality between series x and y at significance levels of 0.05.}
\label{tab:gcoverview}
\scalebox{0.7}{
\begin{tabular}{@{}lllllll@{}}
\toprule
\textbf{x}      & \textbf{y}       & \textbf{Lag 1}         & \textbf{Lag 2}         & \textbf{Lag 3} & \textbf{Lag 4} & \textbf{Lag 5} \\ \midrule
Negative News   & \% Price change  & \textrightarrow           & \textrightarrow           & \textrightarrow   & \textrightarrow   & \textrightarrow   \\
Positive News   & \% Price change  &                        &                        &                &                &                \\
Negative Reddit & \% Price change  & \textleftarrow              & \textrightarrow \textleftarrow & \textleftarrow      & \textleftarrow      & \textleftarrow      \\
Positive Reddit & \% Price change  & \textrightarrow \textleftarrow & \textrightarrow \textleftarrow & \textleftarrow      & \textleftarrow      &                \\
Negative forum  & \% Price change  &                        & \textrightarrow           & \textrightarrow   & \textrightarrow   & \textrightarrow   \\
Positive forum  & \% Price change  &                        &                        &                &                &                \\
Negative news   & \% Volume change &                        & \textrightarrow           &                &                &                \\
Positive news   & \% Volume change & \textrightarrow \textleftarrow & \textrightarrow \textleftarrow &                &                &                \\
Negative Reddit & \% Volume change &                        & \textleftarrow              & \textleftarrow      & \textleftarrow      & \textleftarrow      \\
PositiveReddit  & \% Volume change &                        &                        &                & \textleftarrow      & \textleftarrow      \\
Negative forum  & \% Volume change & \textleftarrow              & \textleftarrow              & \textleftarrow      & \textleftarrow      & \textleftarrow      \\
Positive forum  & \% Volume change & \textrightarrow \textleftarrow & \textleftarrow              & \textleftarrow      & \textleftarrow      & \textleftarrow      \\ \bottomrule
\end{tabular}}
\end{table}

\section{Conclusion and further research}\label{sec:conclusion}
This research provides evidence of the value of including online discourse sentiment originating from various sources in predicting daily Bitcoin exchange movements in price and volume. In parallel, it confirms earlier findings by~\cite{Kaminski2014} that suggest social chatter mirrors preceding price and volume changes that occur on the exchanges. As we now find, this predictive causal relationship is not uni-directional for all channels and should be further dissected by channel, market metric, and sentiment. In line with the findings by~\cite{Kaminski2014}, exchange movements do indeed seem to serve as a predictive indicator for the expressed sentiment on the examined social channels (forum and Reddit) for both volume and price. On the other hand, we do find predictive causality for both negative forum and news sentiment in relation to the average daily exchange price. This becomes apparent from the performed Granger causality test, in which we most notably find that negative news and forum sentiment leads price movements, but price and volume movements lead sentiment on forum and Reddit channels at significance levels of 0.05 or below.  

This relationship can likely be strengthened by improving the classifier accuracies beyond the simple logistic regression and multinomial Naive Bayes models used to reach these results. Feature engineering might also help, as there did not seem to be a clear set of separating features for positive sentiment documents. The true test is the application of these findings into algorithmic Bitcoin trading strategies to determine whether these findings hold up in practice. 

Although the sentiment time series are not included into trading algorithms, the performed tests are sufficient to confirm the hypothesis. As the Granger causality test inherently measures if there is any added value of another time series y (in this case sentiment) in predicting series x,  we can conclude that if this sentiment data were indeed included in a trading model, it has the potential to improve the prediction accuracy. 

The main limitations of the conducted study consist of the relatively small and sparse data set and time frame over which the study was carried out. Collecting sentiment from a wider set of sources will make it possible to get a better sense of the overall sentiment amongst the Bitcoin community and see how sentiment changes over time. Single-year Bitcoin exchange prices are generally not stationary and follow a trend.  Language and geographical bias also play a role. Bitcoin is for example heavily used in China. It then follows that the Chinese Bitcoin market participants will likely also be influenced by online discussions that were not included in this research.  

Secondary news and discussions that do not mention Bitcoin directly were also not investigated in this research, although such news can carry important information concerning Bitcoin. News regarding foreign monetary policy for example could very well have a positive effect on Bitcoin, in that more stringent monetary controls can be circumvented by the use of Bitcoin, driving further demand for the cryptocurrency. For the conducted research, the assumption was made that if this information is truly relevant to Bitcoin, it will spill over into Bitcoin communities. This direct analysis and synthesis of secondary news and discussion in relation to Bitcoin would however bring a speed advantage if applied in trading algorithms.


%
\bibliography{references}

\pagebreak
\pagebreak
\cleardoublepage

\appendix
\section{Appendix}
The code for the conducted experiments can be found at \url{https://github.com/marvinkennis/BitcoinThesis}
\\
\\
The scrapers used for data collection have been open-sourced and documented, and can be found at \url{https://github.com/marvinkennis/BitCollect}
\clearpage
\section{Appendix B}

\begin{table}[]
\centering
\caption{Confusion matrix for logistic regression on news articles}
\label{tab:logregnewscm}
\begin{tabular}{@{}lll@{}}
\toprule
\textsubscript{Pred.} / \textsuperscript{Real} & \textbf{P} & \textbf{N} \\ \midrule
\textbf{P}           & 191  & 114  \\
\textbf{N }          & 62  & 453  \\ \bottomrule
\end{tabular}
\end{table}

\begin{table}[]
\centering
\caption{Confusion matrix for logistic regression on Reddit posts}
\label{tab:logregredditcm}
\begin{tabular}{@{}lll@{}}
\toprule
\textsubscript{Pred.} / \textsuperscript{Real} & \textbf{P} & \textbf{N} \\ \midrule
\textbf{P}           & 208  & 281  \\
\textbf{N }          & 112 & 867  \\ \bottomrule
\end{tabular}
\end{table}

\begin{table}[]
\centering
\caption{Confusion matrix for logistic regression on forum topics}
\label{tab:logregforumcm}
\begin{tabular}{@{}lll@{}}
\toprule
\textsubscript{Pred.} / \textsuperscript{Real} & \textbf{P} & \textbf{N} \\ \midrule
\textbf{P}           & 391  & 303  \\
\textbf{N }          & 209  & 1.048  \\ \bottomrule
\end{tabular}
\end{table}

\begin{table}[]
\centering
\caption{Confusion matrix for logistic regression on IRC Chatter}
\label{tab:logregirccm}
\begin{tabular}{@{}lll@{}}
\toprule
\textsubscript{Pred.} / \textsuperscript{Real} & \textbf{P} & \textbf{N} \\ \midrule
\textbf{P}           & 371  & 318  \\
\textbf{N }          & 159  & 692  \\ \bottomrule
\end{tabular}
\end{table}

\begin{table}[]
\centering
\caption{Confusion matrix for multinomial NB on news articles}
\label{tab:mnbnewscm}
\begin{tabular}{@{}lll@{}}
\toprule
\textsubscript{Pred.} / \textsuperscript{Real} & \textbf{P} & \textbf{N} \\ \midrule
\textbf{P}           & 222  & 83  \\
\textbf{N }          & 66  & 449  \\ \bottomrule
\end{tabular}
\end{table}

\begin{table}[]
\centering
\caption{Confusion matrix for multinomial NB on Reddit posts}
\label{tab:mnbredditcm}
\begin{tabular}{@{}lll@{}}
\toprule
\textsubscript{Pred.} / \textsuperscript{Real} & \textbf{P} & \textbf{N} \\ \midrule
\textbf{P}           & 267  & 222  \\
\textbf{N }          & 176  & 803  \\ \bottomrule
\end{tabular}
\end{table}

\begin{table}[]
\centering
\caption{Confusion matrix for multinomial NB on forum topics}
\label{tab:mnbforumcm}
\begin{tabular}{@{}lll@{}}
\toprule
\textsubscript{Pred.} / \textsuperscript{Real} & \textbf{P} & \textbf{N} \\ \midrule
\textbf{P}           & 405  & 289  \\
\textbf{N }          & 287  & 970  \\ \bottomrule
\end{tabular}
\end{table}

\begin{table}[]
\centering
\caption{Confusion matrix for multinomial NB on IRC Chatter}
\label{tab:mnbirccm}
\begin{tabular}{@{}lll@{}}
\toprule
\textsubscript{Pred.} / \textsuperscript{Real} & \textbf{P} & \textbf{N} \\ \midrule
\textbf{P}           & 393  & 296  \\
\textbf{N }          & 238  & 613  \\ \bottomrule
\end{tabular}
\end{table}

\begin{table*}[]
\centering
\caption{Correlation table of positive sentiment from source at time \textit{t} with metric at\textit{ t+n}. Cells show Pearson correlation coefficient (r) and p-value (p)}
\label{tab:poscor}
\scalebox{0.6}{
%
}
\end{table*}


\begin{table*}[]
\centering
\caption{Correlation table of negative sentiment from source at time \textit{t} with metric at\textit{ t+n}. Cells show Pearson correlation coefficient (r) and p-value (p)}
\label{tab:negcor}
\scalebox{0.6}{
%
}
\end{table*}


\begin{table*}[]
\centering
\caption{Correlation table of cumulative (negative + positive) sentiment from source at time \textit{t} with metric at\textit{ t+n}. Cells show Pearson correlation coefficient (r) and p-value (p)}
\label{tab:cmlcor}
\scalebox{0.6}{
%
}
\end{table*}


\begin{table*}[]
\centering
\caption{F-statistics and probability values for each null hypothesis across different lag-lengths on news sentiment.}
\label{tab:gcnews}
\scalebox{0.6}{
\begin{tabular}{@{}lllllllllll@{}}
\toprule
\textbf{News}                                          & \textbf{Lag: 1} & \textbf{}       & \textbf{Lag: 2} & \textbf{}       & \textbf{Lag: 3} & \textbf{}       & \textbf{Lag: 4} & \textbf{}       & \textbf{Lag: 5} &                 \\ \midrule
\rowcolor[HTML]{FFFFFF} 
Null Hypothesis:                                       & F -Statistic    & Prob.           & F-Statistic     & Prob.           & F-Statistic     & Prob.           & F-Statistic     & Prob.           & F-Statistic     & Prob.           \\ \midrule
\rowcolor[HTML]{FFFFFF} 
CUMULATIVESENT does not Granger Cause ABSCHANGEAVG     & 0.26389         & 0.6078          & 1.53485         & 0.2169          & 4.86419         & \textbf{0.0025} & 3.59847         & \textbf{0.0068} & 2.82964         & \textbf{0.0161} \\ \midrule
\rowcolor[HTML]{FFFFFF} 
ABSCHANGEAVG does not Granger Cause CUMULATIVESENT     & 0.0169          & 0.8966          & 0.29422         & 0.7453          & 0.46331         & 0.7081          & 0.3604          & 0.8367          & 0.43122         & 0.8268          \\
\rowcolor[HTML]{FFFFFF} 
                                                       &                 &                 &                 &                 &                 &                 &                 &                 &                 &                 \\
\rowcolor[HTML]{FFFFFF} 
NEGATIVESENT does not Granger Cause ABSCHANGEAVG       & 3.6834          & 0.0557          & 3.79035         & \textbf{0.0235} & 3.46922         & \textbf{0.0164} & 3.06871         & \textbf{0.0166} & 2.84926         & \textbf{0.0155} \\ \midrule
\rowcolor[HTML]{FFFFFF} 
ABSCHANGEAVG does not Granger Cause NEGATIVESENT       & 0.01306         & 0.9091          & 0.0114          & 0.9887          & 0.00315         & 0.9998          & 1.51E-01        & 9.62E-01        & 0.14615         & 0.98            \\
\bottomrule
\end{tabular}}
\end{table*}


\begin{table*}[]
\centering
\caption{F-statistics and probability values for each null hypothesis across different lag-lengths on Reddit sentiment.}
\label{tab:gcreddit}
\scalebox{0.6}{
\begin{tabular}{@{}lllllllllll@{}}
\toprule
\textbf{Reddit}                                        & \textbf{Lag 1} & \textbf{}       & \textbf{Lag 2} & \textbf{}                  & \textbf{Lag 3} & \textbf{}                  & \textbf{Lag 4} & \textbf{}                  & \textbf{Lag 5} & \textbf{}                  \\ \midrule
Null Hypothesis:                                       & F-Statistic    & Prob.           & F-Statistic    & Prob.                      & F-Statistic    & Prob.                      & F-Statistic    & Prob.                      & F-Statistic    & Prob.                      \\ \midrule
ABSCHANGEAVG does not Granger Cause CUMULATIVESENT     & 0.70582        & 0.4014          & 0.52859        & 0.5899                     & 0.48504        & 0.6929                     & 0.43334        & 0.7845                     & 0.45817        & 0.8073                     \\ \midrule
CUMULATIVESENT does not Granger Cause ABSCHANGEAVG     & 6.39111        & \textbf{0.0119} & 4.4292         & \textbf{0.0126}            & 3.21015        & \textbf{0.0232}            & 2.41033        & \textbf{0.049}             & 2.59565        & \textbf{0.0254}            \\
                                                       &                &                 &                &                            &                &                            &                &                            &                &                            \\
ABSVOLUMECHANGE does not Granger Cause CUMULATIVESENT  & 0.85451        & 0.3559          & 3.18711        & \textbf{0.0425}            & 2.83765        & \textbf{0.038}             & 2.11369        & 0.0787                     & 1.82367        & 0.1075                     \\
\bottomrule
\end{tabular}}
\end{table*}

\begin{table*}[]
\centering
\caption{F-statistics and probability values for each null hypothesis across different lag-lengths on forum sentiment.}
\label{tab:gcforum}
\scalebox{0.6}{
\begin{tabular}{@{}lllllllllll@{}}
\toprule
\textbf{Forum}                                         & \textbf{Lags: 1} & \textbf{}                   & \textbf{Lags: 2} & \textbf{}                   & \textbf{Lags: 3} & \textbf{}                   & \textbf{Lags: 4} & \textbf{}       & \textbf{Lags: 5} & \textbf{}       \\ \midrule
Null Hypothesis:                                       & F-Statistic      & Prob.                       & F-Statistic      & Prob.                       & F-Statistic      & Prob.                       & F-Statistic      & Prob.           & F-Statistic      & Prob.           \\ \midrule
CUMULATIVESENT does not Granger Cause ABSCHANGEAVG     & 5.37386          & \textbf{0.021}              & 4.78089          & \textbf{0.0089}             & 2.97202          & \textbf{0.0318}             & 2.49198          & \textbf{0.0429} & 1.99967          & 0.0781          \\ \midrule
ABSCHANGEAVG does not Granger Cause CUMULATIVESENT     & 12.6036          & \textbf{0.0004}             & 6.21084          & \textbf{0.0022}             & 4.24911          & \textbf{0.0057}             & 3.06509          & \textbf{0.0167} & 2.30655          & \textbf{0.0441} \\
                                                       &                  &                             &                  &                             &                  &                             &                  &                 &                  &                 \\
NEGATIVESENT does not Granger Cause ABSCHANGEAVG       & 3.87527          & \textbf{0.0498}             & 3.91498          & \textbf{0.0208}             & 3.9053           & \textbf{0.0091}             & 3.45839          & \textbf{0.0087} & 3.09085          & \textbf{0.0096} \\
\bottomrule
\end{tabular}}
\end{table*}

\end{document}